\documentclass[baaa]{baaa}
\def\msol{M$_\odot$}
 
\usepackage[pdftex]{hyperref}
\usepackage{subfigure}
\usepackage{natbib}
\usepackage{helvet,soul}
\usepackage[font=small]{caption}
\usepackage{graphicx}

\contriblanguage{1}

\contribtype{1}

\thematicarea{99}

\received{\ldots}
\accepted{\ldots}

\title{Characterization of molecular outflows at core-scale in the massive clump AGAL G345.0029$-$0.224}

\titlerunning{Characterization of molecular outflows at core-scale in the massive clump AGAL G345.0029$-$0.224}

\author{E. Cohen Arazi\inst{1,2}, M. Ortega\inst{1},  S. Paron\inst{1}, P. F.  Velázquez\inst{3}, A. Rodríguez-González\inst{3}, \& E. Alquicira\inst{3}}

\authorrunning{Cohen Arazi et al.}

\contact{eitan1174@gmail.com}

\institute{
Instituto de Astronomía y Física del Espacio, CONICET-UBA, Argentina
\and
Departamento de Física, Facultad de Ciencias Exactas y Naturales, UBA, Argentina
\and
Universidad Nacional Autónoma de México, Instituto de Ciencias Nucleares, A.P. 70-543, 04510 Ciudad de México, México
}

\resumen{Las estrellas de alta masa con sus fuertes vientos e intensos campos de radiación son fundamentales en la regulación de la dinámica y la evolución galáctica; sin embargo, a pesar de su gran relevancia aún no se comprenden en profundidad los mecanismos involucrados en su formación. En este contexto, los flujos salientes (outflows) moleculares esenciales para remover el momento angular y permitir la acreción hacia el objeto central, son un fenómeno crucial a la hora de caracterizar su formación. Estudios previos revelan una discrepancia en las masas de los \textit{outflows} asociados a grumos de alta masa entre trabajos llevados a cabo a escala de grumo ($\sim$ pc) y a escala de núcleo ($\sim$ subpc). Esto sugiere que la actividad de \textit{outflows} de alta masa observada a escala de grumo podría ser quizás el resultado de la contribución de varios \textit{outflows} de menor masa vinculados a núcleos moleculares individuales. Este trabajo presenta un estudio del gas molecular hacia un grumo de alta masa asociado con un Objeto Verde Extendido (EGO, según su sigla en inglés). Los EGOs son indicadores de chorros (jets) asociados a protoestrellas de alta masa. Empleando datos de alta resolución angular del Atacama Large Millimeter/submillimeter Array  se observó en la fuente la presencia de varios núcleos calientes con actividad de \textit{outflow}. Se presenta una caracterización de los \textit{outflows} a escala de núcleo en el contexto de los parámetros físicos del grumo molecular.}

\abstract{High-mass stars, with their powerful winds and intense radiation fields, are fundamental in regulating galactic dynamics and evolution; however, despite their great relevance, the mechanisms involved in their formation are still not fully understood. In this context, molecular outflows, which are essential for removing angular momentum and allowing accretion onto the central object, are a crucial phenomenon for characterizing their formation.
Previous studies reveal a discrepancy in the masses of outflows associated with high-mass clumps between works conducted at the clump scale ($\sim$ pc) and those at the core scale ($\sim$ subpc). This suggests that the high-mass outflow activity observed at the clump scale might be the result of the contribution from several lower-mass outflows linked to individual molecular cores.
This work presents a study of the molecular gas toward a high-mass clump associated with an Extended Green Object (EGO). EGOs are indicators of jets associated with high-mass protostars. Employing high angular resolution data from the Atacama Large Millimeter/submillimeter Array (ALMA), the presence of several hot cores with outflow activity was observed in the source. A characterization of the outflows at the core scale is presented within the context of the physical parameters of the molecular clumps.}


\begin{document}

\maketitle
\section{Introduction}\label{S_intro}

A ubiquitous phenomenon occurring during the formation of a star is the generation of molecular outflows. They are produced by the ambient molecular material that is pushed out by jets ejected at the protostar-disk scale \citep{bally2016,bj16}. Such jets and molecular outflows reduce the angular momentum of the protostar-accretion disk system, allowing the central object to accumulate material.

There are some statistical studies that characterize molecular outflows at the clump scale in samples of high-mass molecular clumps using single\textbf{-}dish observations \citep[see][]{maud2015,yang2018}. These studies show that the physical properties of high-mass outflows, such as mass, energy, luminosity, etc., are closely related to the properties of the molecular clump that hosts them. In particular, it is usually estimated a mass range for high-mass outflows from 1 \msol~to $10^3$ \msol, several orders of magnitude greater than the masses of low-mass outflows (about 0.1 \msol). 

\begin{figure*}[t]
    \centering
    \includegraphics[width=0.4\textwidth]{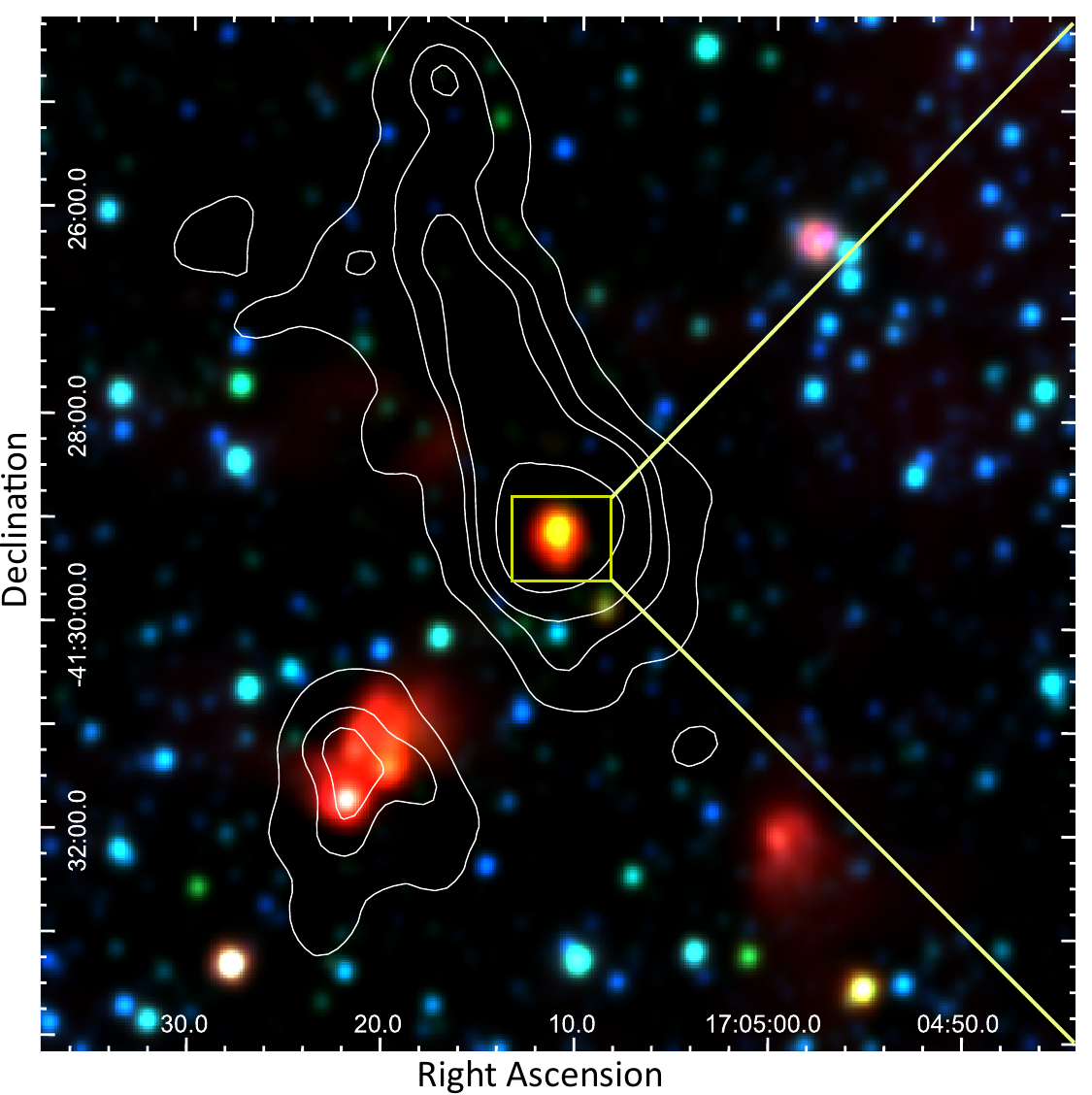}
    \includegraphics[width=0.4\textwidth]{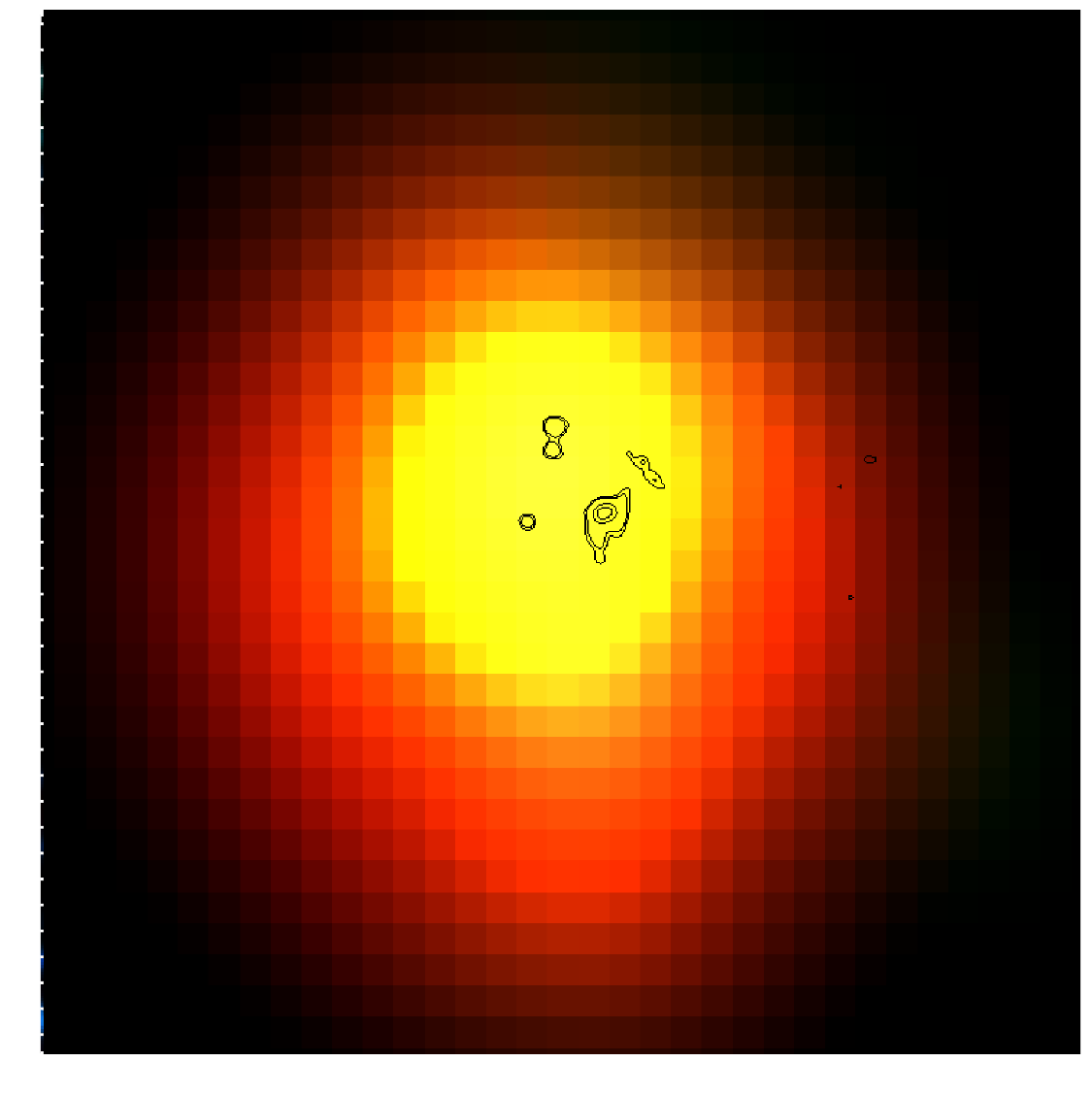}
    \caption{\textit{The left panel} shows a WISE three-colour composite image, with 3.4 $\mu$m in blue, 4.6 $\mu$m in green, and 12 $\mu$m in red. The white contours represent the 870 $\mu$m emission from ATLASGAL. Levels are at 300, 600, 900, 1300 mJy beam$^{-1}$. \textit{The right panel} is a close-up view of the ATLASGAL source 345.0029$-$0.2241 in which EGO G345.00$-$0.22(a) is embedded (yellow area). The black contours represent the ALMA continuum emission at 340 GHz. Levels are at 50, 100, 300, 700 mJy beam$^{-1}$.}
    \label{fig:wise_alma}
\end{figure*}

Some years ago, \citet{li2020} conducted a study on molecular outflows associated with cores embedded in high-mass clumps at early evolutionary stages (infrared-quiet sources). Using high-angular resolution observations from ALMA, they identified outflows with significantly lower masses compared to previous studies, ranging from 0.001 to 0.32 \msol. The research revealed a strong correlation between some physical parameters, such as the total mechanical force of the outflows (F$_{\rm out}$) in a given clump and the bolometric luminosity-mass ratio of the clump (related to the evolutionary stage). Additionally, the authors found that in high-mass star forming regions, outflow activity seemed to intensify as the sources evolved. 

Aiming to characterize molecular outflow activity at  core spatial scales, we present a high-angular resolution study of the molecular environment associated with EGO G345.00$-$0.22(a). EGO stands for Extended Green Object, since they are sources that present extended emission in the 4.5 $\mu$m band, which by convention are usually displayed in that color. The particularity of this mid-infrared band is that it can contain some emission lines of $\text{H}_2$ that are only excited under shock conditions, suggesting the possible presence of massive molecular outflows.

EGO G345.00-0.22(a), catalogued by \citet{cyganowski2008}, is embedded in the massive clump AGAL 345.0029$-$0.2241 (hereafter G345), where AGAL refers to the ATLASGAL (Atacama Pathfinder EXperiment (APEX) Telescope Large Area Survey of the Galaxy) region type. \citet{whitaker2017} determined a v$_{\rm LSR}$ of $-$29.3 km s$^{-1}$ and estimated a kinematical distance of about 3.1 kpc for the molecular clump. \citet{bronfman1996} catalogued the source as an UC H\textsc{ii}~region, and \citet{konig2017} estimated a bolometric luminosity and a mass of $6 \times 10^4$ L$_\odot$ and $1 \times 10^3$ \msol, respectively, for the clump. 

\section{Data}

\subsection{ALMA data}

Data cubes were obtained from the ALMA Science Archive \footnote{https://almascience.nrao.edu/aq/}. We used data from the 2017.1.00914 project (PI: Csengeri, T.). The telescope configuration used baselines  L5BL/L80BL of 34.9/229.2 m, respectively,  in the 12\,m array. 
The observed frequency range of the spectral window used in this work spans from 333.461 to 349.086~GHz (Band 7). The angular and spectral resolutions are 0$.\!\!^{\prime\prime}$45 ($\approx$ 0.006 pc at 3.1 kpc) and 1.1~MHz, respectively. The velocity resolution is about 0.9~km s$^{-1}$. The root mean square (rms) noise level is 3.7~mJy~beam$^{-1}$ for the emission line (averaged each 10~km s$^{-1}$) and 0.15~mJy~beam$^{-1}$ for the continuum emission. The maximum recoverable spatial scale is $6.5^{\prime\prime}$. 

\subsection{Additional data}

\begin{itemize}

\item An image of ATLASGAL at 870 $\mu$m (345 GHz)  \citep{schuller2009} was used. The images have a rms noise in the range of 0.05--0.07 Jy beam$^{-1}$ and a beam size of $\approx$ 19 $\!\!^{\prime\prime}$.  
 
\item We also used images at 3.4, 4.6, and 12 $\mu$m extracted from the  Wide-field Infrared Survey Explorer (WISE; \citealt{wright2010}). The angular resolution and sensitivity are about 6 $\!\!^{\prime\prime}$ and 0.1 mJy, respectively.

\end{itemize}

\section{Results}

Figure\,\ref{fig:wise_alma}, left panel shows a WISE three-color composite image, with the 3.4, 4.6, and 12 $\mu$m emissions displayed in blue, green, and red, respectively. The white contours represent the 870 $\mu$m emission from ATLASGAL. Figure\,\ref{fig:wise_alma},right panel is a close-up view of G345 in which the EGO is embedded. The black contours represent the ALMA continuum emission at 340 GHz. The molecular clump, traced by the maximum of the ATLASGAL emission, exhibits a clear fragmentation with several molecular cores as observed from the ALMA continuum emission.

\begin{figure}[h]
    \centering
    \includegraphics[width=6cm]{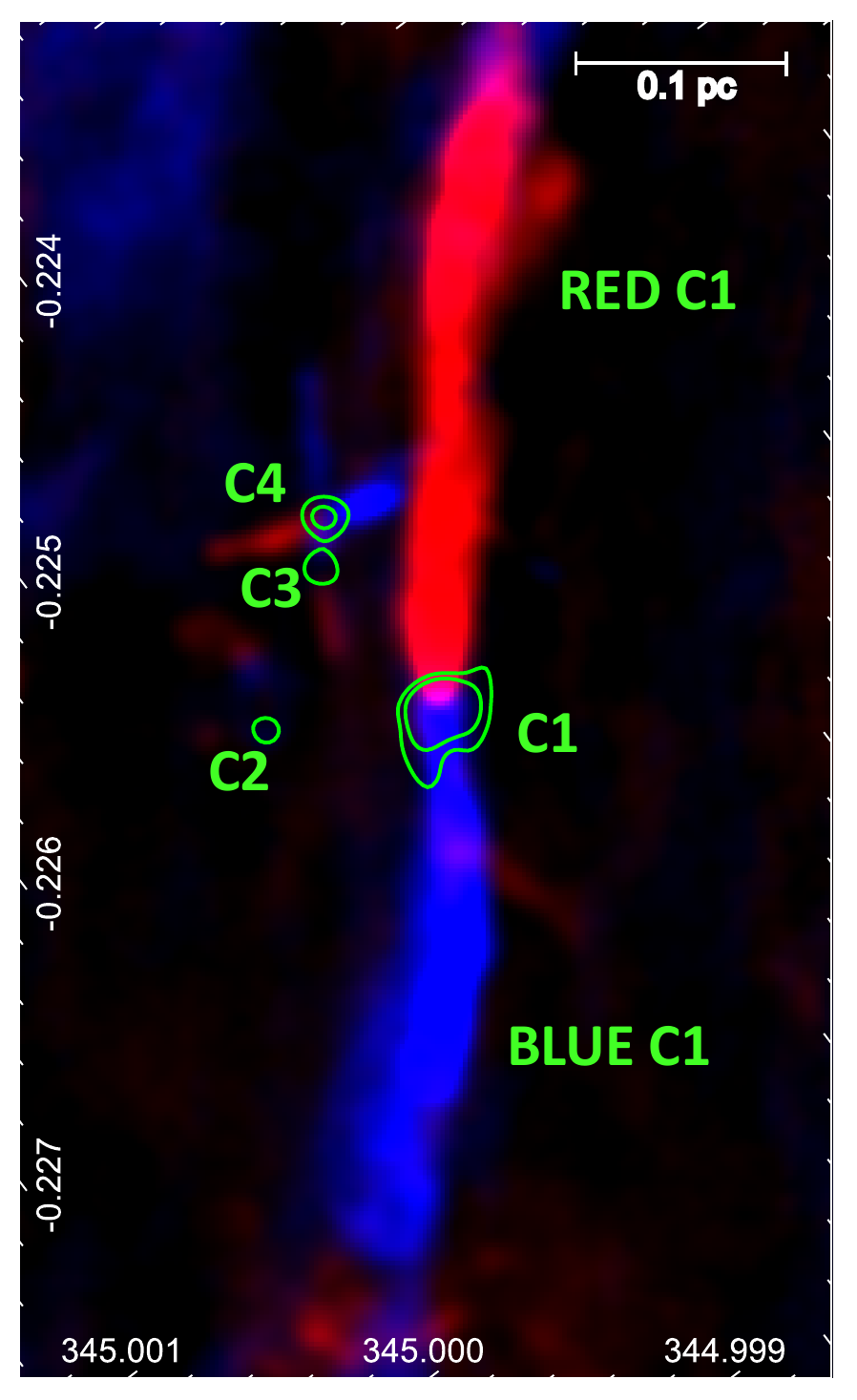}
    \caption{$^{12}$CO J=3--2 emission distribution integrated (moment 0) between $-60$ and $-10$ km s$^{-1}$(blue), and between $+5$ and $+65$ km s$^{-1}$(red) considering the $^{12}$CO J=3--2 transition at rest (0 km s$^{-1}$). The continuum emission at 340 GHz is represented in green contours with levels of 100 and 200 mJy beam$^{-1}$}
    \label{fig:outflows_mom0}
\end{figure}

In order to study the possible presence of molecular outflows, we analyze the  $^{12}$CO emission from the data cube. Figure\,\ref{fig:outflows_mom0} shows the $^{12}$CO J=3--2 emission distribution integrated between $-60$ and $-10$ km s$^{-1}$ (blue), and between $+5$ and $+65$ km s$^{-1}$ (red) towards the region in which the discovered cores (in green contours) lie. 
The most conspicuous molecular core, labelled C1, exhibits the most noticeable molecular outflow activity with well-collimated lobes aligned in the north-south direction. C3 and C4 cores also show molecular outflow activity, although less intense, with lobes aligned perpendicular to each other. The C2 core does not have detected outflow activity. 

To characterize the molecular outflows observed in the region, we assume local thermodynamic equilibrium (LTE) conditions and an optically thin regime for the $^{12}$CO J=3--2 spectral wings. Following the procedures detailed in \citet{ortega2023}, we obtained the outflow parameters presented in Table\,\ref{tab:outflow_params}.

\begin{figure}
    \centering
    \includegraphics[width=8cm]{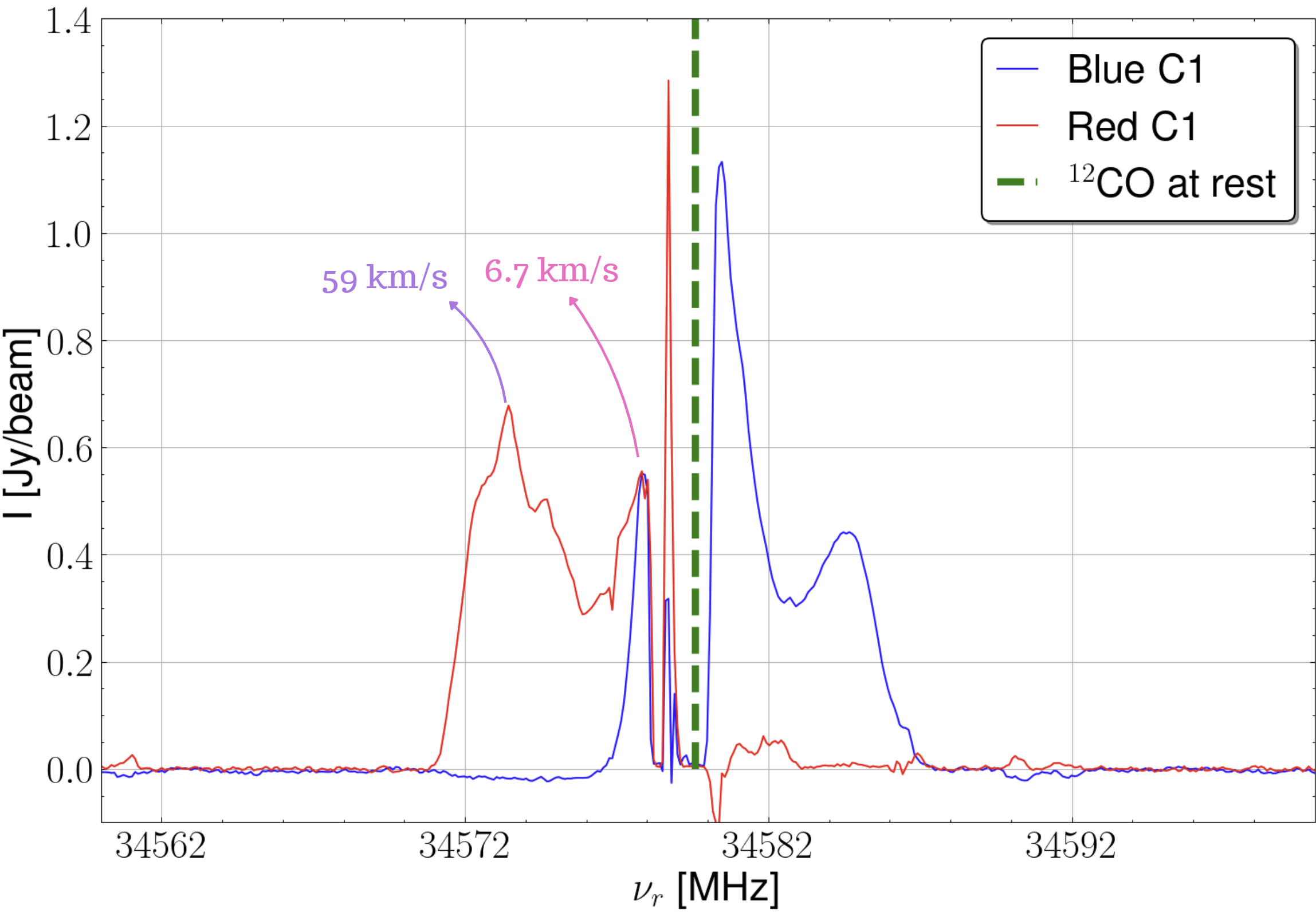}
    \caption{$^{12}$CO J=3--2 spectra obtained towards the red- and blue-shifted lobes (red and blue curves respectively) related to core C1. Both lobes exhibit high and low velocity components, which are labeled in the red lobe. The green dashed line indicates the rest frequency of the $^{12}$CO J=3--2 transition.}
    \label{fig:spectra}
\end{figure}

\begin{figure}[t]
    \centering
    \includegraphics[width=4cm]{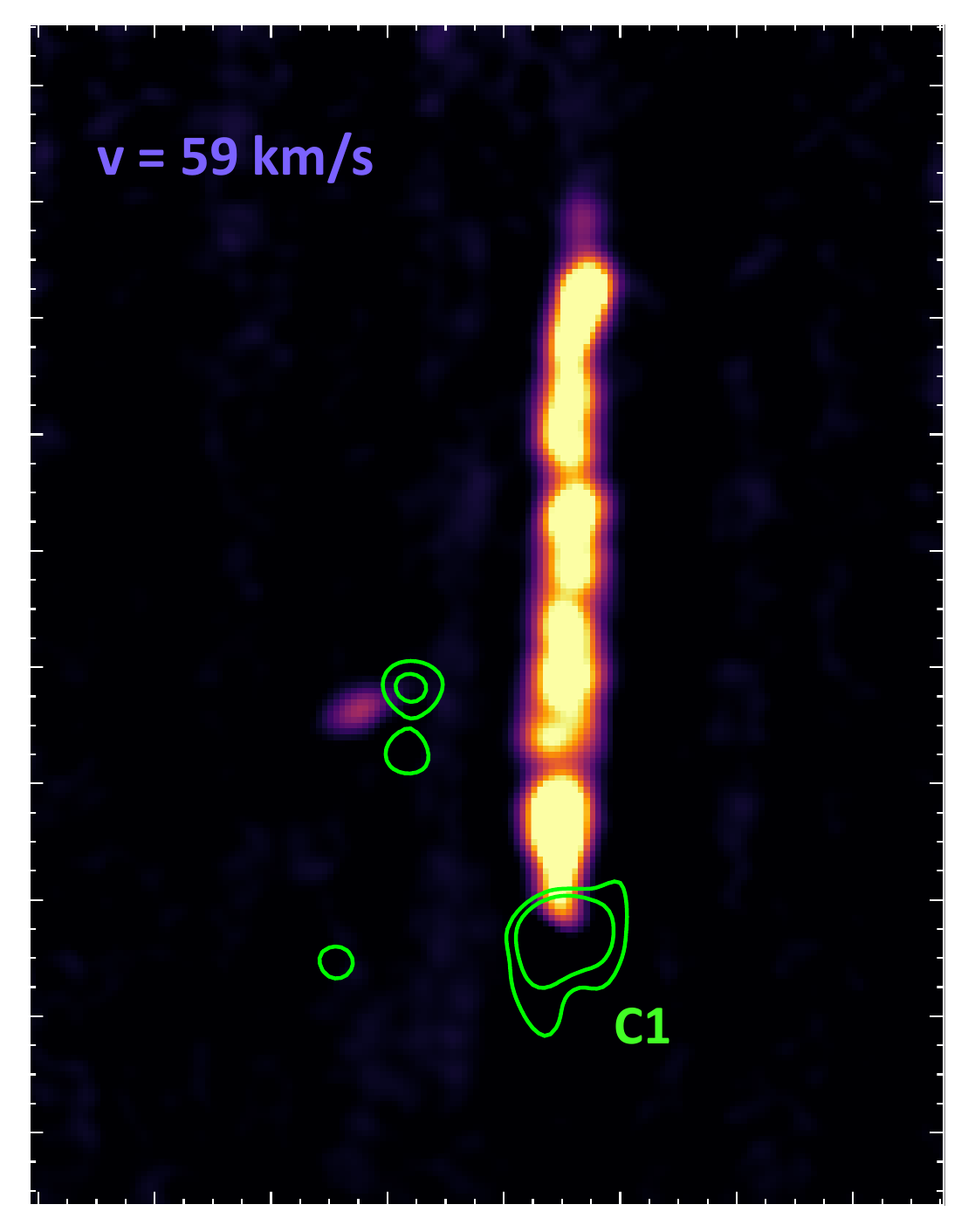}
    \includegraphics[width=4cm]{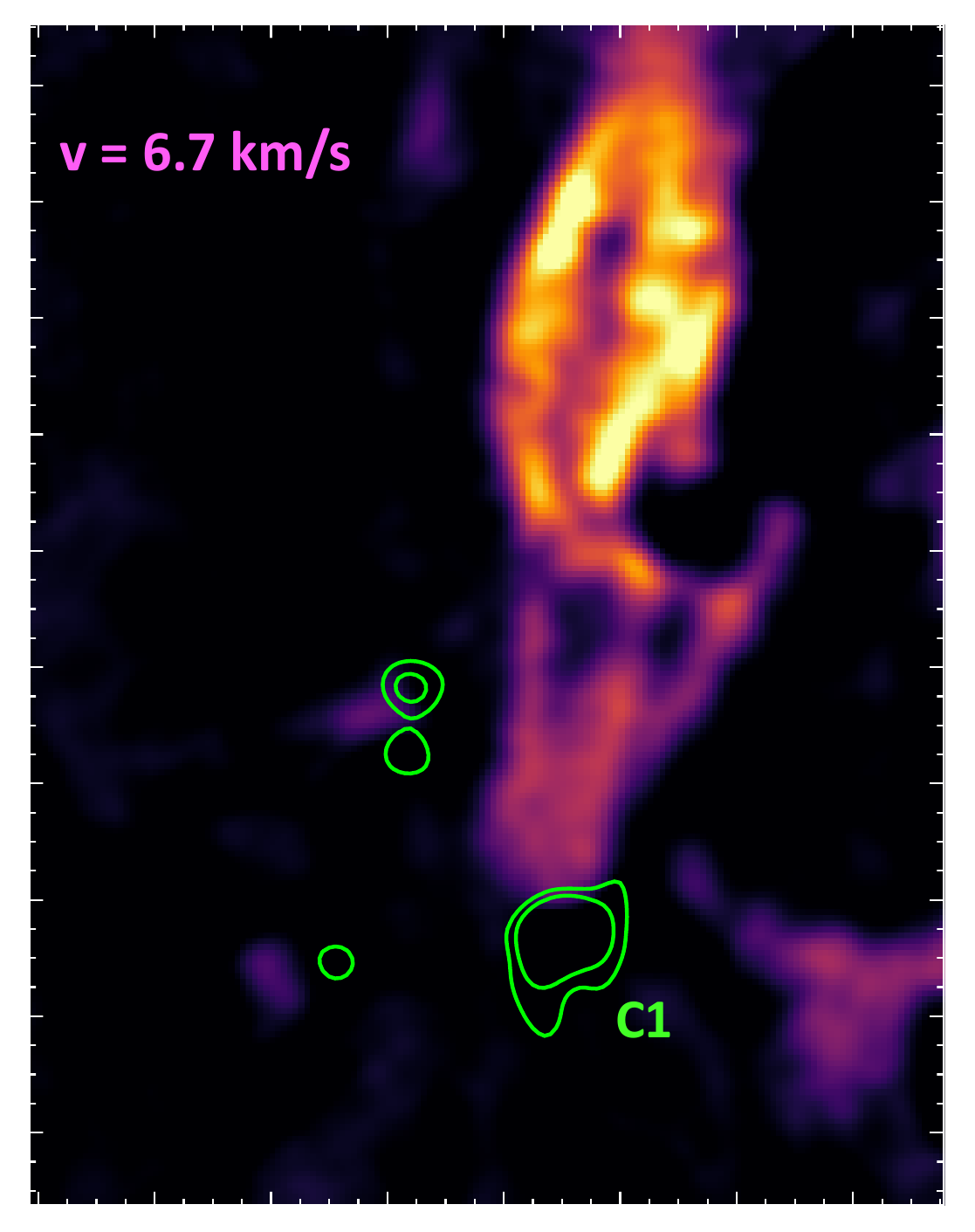}
    \caption{High- and low-velocity molecular outflow towards the core C1. \textit{Left panel}: High-velocity outflow at 59 km s$^{-1}$ of the $^{12}$CO J=3--2 transition. \textit{Right panel}: Low-velocity outflow at 6.7 km s$^{-1}$ of the $^{12}$CO J=3--2 transition. Both velocity planes are highlighted in the $^{12}$CO J=3--2 spectrum of Fig. \ref{fig:spectra}. The green contours represent the ALMA continuum emission at 340 GHz. Levels are at 100 and 200 mJy beam$^{-1}$. }
    \label{fig:velocidades}
\end{figure}

Figure\,\ref{fig:spectra} shows the superposition of the spectra taken towards the red and blue lobes of the molecular outflow related to core C1. The green dashed line indicates the rest frequency of the $^{12}$CO J=3--2 transition. It can be clearly seen that both red and blue wings have  high- and low-velocity components. In particular, a component at 6.7 km s$^{-1}$ (low-velocity) and other one at 59 km s$^{-1}$ (high-velocity) are indicated in the red spectral wing. These molecular components are shown in the maps presented in
Fig.\,\ref{fig:velocidades}. The $^{12}$CO emission channels of the high- and low-velocity components are displayed in the left and right panels, respectively. It is worth noting that the high-velocity gas exhibits a well-collimated structure likely tracing the internal layers of gas in the outflow lobe, while the low-velocity gas shows a less-collimated morphology, tracing the external layers of the analyzed outflow.

\begin{table*}[h!]
\centering
\caption{Main parameters of the C1, C3 and C4 molecular outflows.}
\label{tab:outflow_params}
\resizebox{\textwidth}{!}{%
\begin{tabular}{lcccccccc}
\hline\hline
Parameter & Red-C1 & Blue-C1 & Red-C3 & Blue-C3 & Red-C4 & Blue-OC4 \\
\hline
Mass ($\rm M_\odot$) & $0.388 \pm 0.039$ & $0.548 \pm 0.061$ & $0.004 \pm 0.001$ & $0.007 \pm 0.001$ & $0.009 \pm 0.001$ & $0.021 \pm 0.002$ \\
Momentum ($\rm M_\odot$\,km\,s$^{-1}$) & $13.47 \pm 1.46$ & $9.95 \pm 1.05$ & $0.06 \pm 0.01$ & $0.15 \pm 0.02$ & $0.18 \pm 0.03$ & $0.46 \pm 0.06$ \\
Energy ($10^{45}$\,erg) & $4.65\pm0.61$ & $1.79\pm0.22$ & $(9.83 \pm 1.19)\times10^{-3}$ & $(3.16\pm 0.54)\times10^{-2}$ & $(3.58\pm 0.81)\times10^{-2}$ & $(9.97\pm0.21)\times10^{-2}$ \\
$\rm F_{\mathrm{out}}$ ($\times10^{-3}\,\rm M_\odot$\,km\,s$^{-1}$\,yr$^{-1}$) & $4.90\pm0.67$ & $2.94\pm0.43$ & $(4.36\pm 0.61)\times10^{-2}$  &  $(1.29\pm 0.19)\times10^{-1}$ & $(3.75\pm 0.61)\times10^{-1}$ & $(7.70\pm0.12)\times10^{-1}$ \\
\hline
Length (pc) & $0.22 \pm 0.02$ & $0.25 \pm 0.02$ & $0.032 \pm 0.007$ & $0.051 \pm 0.006$ & $0.033 \pm 0.006$ & $0.043 \pm 0.006$ \\
Dynamical age (yr) & $2752 \pm 257$ & $3388 \pm 275$ & $1344 \pm 120$ & $1147 \pm 96$ & $480 \pm 39$ & $592 \pm 48$ \\
\hline
\end{tabular}%
}
\end{table*}

\section{Discussion}

This section is primarily intended to analyze the core-scale molecular outflow parameters obtained in this work (see Table\,\ref{tab:outflow_params}) aiming to compare them with characterizations done at the molecular clump spatial scales found in the literature.

\citet{guerra2023} characterized the molecular outflow activity at the clump scale in G345 (named IRAS 17016$-$4124 in their work) using the SiO J=4--3 line  (173.688 GHz) with APEX observations (beam $\approx$ 36$^{\prime \prime}$). Table\,\ref{guerra-outflows} compares the values estimated by \citet{guerra2023} with the sum of the contributions of the blue and red lobes for the core C1 obtained in our work. It can be seen that the values derived from clump spatial scales exceed by more than an order of magnitude those calculated by us at the core spatial scale. In particular, the clump-scale mass of the outflow is about 60 times greater than the core-scale outflow mass. 

Focusing on the energy parameter of the molecular outflow activity, there is no physical explanation for the core-scale energy to increase by a factor of nearly 122 at the clump scale. We hypothesize that this discrepancy between the different scales results from an overestimation of the values in the study carried out with single dish. This is likely because of the low angular resolution of the observations that includes surrounding gas that is not directly related to the small-scale outflow activity.

\begin{table}[h!]
\centering
\caption{Comparison between the clump-scale molecular outflow parameters derived by \citet{guerra2023} (column 2) and the total contribution of the blue and red lobes of core C1 obtained in this work (column 3).}
\label{guerra-outflows}
\begin{tabular}{lcc}
\hline\hline
Outflow parameter & Clump-scale & Core C1 \\
                  & APEX        &  ALMA \\   
\hline
Mass ($\rm M_\odot$) & 55 & 0.9  \\
Momentum ($\rm M_\odot$\,km\,s$^{-1}$) & 2064 & 23.2\\
Energy ($10^{45}$\,erg) &  780 & 6.4\\
$\rm F_{\mathrm{out}}$ ($\times10^{-3}\, \rm M_\odot$\,km\,s$^{-1}$\,yr$^{-1}$) &  320 & 7.8\\
Dynamical age ($\times10^{3}$\,yr) & 6 & 3.1\\
\hline
\end{tabular}
\end{table}

On the other hand, it is interesting to compare our results with those obtained by \citet{li2020}, who characterized  molecular outﬂows at core-scale at early stages of high-mass star formation towards 12 infrared dark cloud (IRDC) clumps. The authors used ALMA observations with an angular resolution of about $1.2^{\prime\prime}$, which is comparable to that of our work.

It is interesting to note that all of our parameters, except the dynamical timescales, are larger than the maximum values estimated by the authors for their sample of 43 studied molecular outflows. This shows that the molecular outflow associated with core C1 is undoubtedly of high mass. Conversely, it is worth mentioning that since the source studied here is more evolved than those analyzed by \citet{li2020}, it is expected that the outflows related to C1 in EGO 345.00-0.22(a) are more massive, given that, as the authors indicated, outflow activity appears to increase as the source evolves.

Finally, \citet{li2020} found a correlation between the total outflow force (F$_{\rm out,tot}$) of all outflows associated with a given clump and the clump luminosity-to-mass ratio (L/M) (see Fig.\,5, left panel of their work). G345, with F$_{\rm out,tot}$ of approximately $8 \times 10^{-3}$ $\rm M_\odot$\,km\,s$^{-1}$\,yr$^{-1}$ and a clump bolometric luminosity-mass ratio of approximately 60 L$_{\odot}$/M$_{\odot}$, is located, as expected, among the more evolved sources in the figure.

\section{Summary and conclusion}

We have presented a characterization of the molecular outflow activity associated with the massive clump G345.0029-0.224. Using high angular resolution ALMA observations, we have detected several molecular cores, three of which exhibit clear outflow activity. The most prominent one, core C1, shows a well-collimated bipolar outflow aligned in the north–south direction.

By comparing our derived parameters with those obtained from single-dish observations at the clump scale, we find that the latter overestimate outflow properties by more than an order of magnitude, likely due to contamination from unrelated large-scale gas. Our results, therefore, reinforce the need for high-resolution studies to accurately determine the physical properties of outflows in massive star-forming regions.

Finally, the physical parameters of the outflow related to core C1 are consistent with those expected for massive protostellar sources and suggest that the outflow activity increases as the source evolves. This work contributes to bridging the gap between clump- and core-scale studies, providing new insight into the role of molecular outflows in high-mass star formation.

\begin{acknowledgement}
This work was partially supported by the Argentine grants PIP 2021 11220200100012 and PICT 2021-GRF-TII-00061 awarded by CONICET and ANPCYT.
\end{acknowledgement}


\bibliographystyle{baaa}
\small
\bibliography{bibliografia}

@ARTICLE{guerra2023,
       author = {{Guerra-Varas}, N. and {Merello}, M. and {Bronfman}, L. and {Duronea}, N. and {Elia}, D. and {Finger}, R. and {Mendoza}, E.},
        title = "{SiO outflows in the most luminous and massive protostellar sources of the southern sky}",
      journal = {\aap},
         year = 2023,
        month = sep,
       volume = {677},
          eid = {A148},
        pages = {A148},
          doi = {10.1051/0004-6361/202245522},
archivePrefix = {arXiv},
       eprint = {2307.16350},
 primaryClass = {astro-ph.GA},
       adsurl = {https://ui.adsabs.harvard.edu/abs/2023A&A...677A.148G}
}

@ARTICLE{ortega2023,
       author = {{Ortega}, M.~E. and {Martinez}, N.~C. and {Paron}, S. and {Marinelli}, A. and {Isequilla}, N.~L.},
        title = "{Looking for evidence of high-mass star formation at core scale in a massive molecular clump}",
      journal = {\aap},
         year = 2023,
        month = sep,
       volume = {677},
          eid = {A129},
        pages = {A129},
          doi = {10.1051/0004-6361/202346661},
archivePrefix = {arXiv},
       eprint = {2307.03644},
 primaryClass = {astro-ph.GA},
       adsurl = {https://ui.adsabs.harvard.edu/abs/2023A&A...677A.129O}
}

@ARTICLE{schuller2009,
       author = {{Schuller}, F. and {Menten}, K.~M. and {Contreras}, Y. and {Wyrowski}, F. and {Schilke}, P. and {Bronfman}, L. and {Henning}, T. and {Walmsley}, C.~M. and {Beuther}, H. and {Bontemps}, S. and {Cesaroni}, R. and {Deharveng}, L. and {Garay}, G. and {Herpin}, F. and {Lefloch}, B. and {Linz}, H. and {Mardones}, D. and {Minier}, V. and {Molinari}, S. and {Motte}, F. and {Nyman}, L. -{\r{A}}. and {Reveret}, V. and {Risacher}, C. and {Russeil}, D. and {Schneider}, N. and {Testi}, L. and {Troost}, T. and {Vasyunina}, T. and {Wienen}, M. and {Zavagno}, A. and {Kovacs}, A. and {Kreysa}, E. and {Siringo}, G. and {Wei{\ss}}, A.},
        title = "{ATLASGAL - The APEX telescope large area survey of the galaxy at 870 {\ensuremath{\mu}}m}",
      journal = {\aap},
         year = 2009,
        month = sep,
       volume = {504},
       number = {2},
        pages = {415-427},
          doi = {10.1051/0004-6361/200811568},
archivePrefix = {arXiv},
       eprint = {0903.1369},
 primaryClass = {astro-ph.GA},
       adsurl = {https://ui.adsabs.harvard.edu/abs/2009A&A...504..415S}
}

@ARTICLE{wright2010,
       author = {{Wright}, Edward L. and {Eisenhardt}, Peter R.~M. and {Mainzer}, Amy K. and {Ressler}, Michael E. and {Cutri}, Roc M. and {Jarrett}, Thomas and {Kirkpatrick}, J. Davy and {Padgett}, Deborah and {McMillan}, Robert S. and {Skrutskie}, Michael and {Stanford}, S.~A. and {Cohen}, Martin and {Walker}, Russell G. and {Mather}, John C. and {Leisawitz}, David and {Gautier}, III, Thomas N. and {McLean}, Ian and {Benford}, Dominic and {Lonsdale}, Carol J. and {Blain}, Andrew and {Mendez}, Bryan and {Irace}, William R. and {Duval}, Valerie and {Liu}, Fengchuan and {Royer}, Don and {Heinrichsen}, Ingolf and {Howard}, Joan and {Shannon}, Mark and {Kendall}, Martha and {Walsh}, Amy L. and {Larsen}, Mark and {Cardon}, Joel G. and {Schick}, Scott and {Schwalm}, Mark and {Abid}, Mohamed and {Fabinsky}, Beth and {Naes}, Larry and {Tsai}, Chao-Wei},
        title = "{The Wide-field Infrared Survey Explorer (WISE): Mission Description and Initial On-orbit Performance}",
      journal = {\aj},
         year = 2010,
        month = dec,
       volume = {140},
       number = {6},
        pages = {1868-1881},
          doi = {10.1088/0004-6256/140/6/1868},
archivePrefix = {arXiv},
       eprint = {1008.0031},
 primaryClass = {astro-ph.IM},
       adsurl = {https://ui.adsabs.harvard.edu/abs/2010AJ....140.1868W}
}

@ARTICLE{konig2017,
       author = {{K{\"o}nig}, C. and {Urquhart}, J.~S. and {Csengeri}, T. and {Leurini}, S. and {Wyrowski}, F. and {Giannetti}, A. and {Wienen}, M. and {Pillai}, T. and {Kauffmann}, J. and {Menten}, K.~M. and {Schuller}, F.},
        title = "{ATLASGAL-selected massive clumps in the inner Galaxy. III. Dust continuum characterization of an evolutionary sample}",
      journal = {\aap},
         year = 2017,
        month = mar,
       volume = {599},
          eid = {A139},
        pages = {A139},
          doi = {10.1051/0004-6361/201526841},
archivePrefix = {arXiv},
       eprint = {1610.09055},
 primaryClass = {astro-ph.GA},
       adsurl = {https://ui.adsabs.harvard.edu/abs/2017A&A...599A.139K}
}

@ARTICLE{bronfman1996,
       author = {{Bronfman}, L. and {Nyman}, L. -A. and {May}, J.},
        title = "{A CS(2-1) survey of IRAS point sources with color characteristics of ultra-compact HII regions.}",
      journal = {\aaps},
         year = 1996,
        month = jan,
       volume = {115},
        pages = {81},
       adsurl = {https://ui.adsabs.harvard.edu/abs/1996A&AS..115...81B}
}

@ARTICLE{whitaker2017,
       author = {{Whitaker}, J. Scott and {Jackson}, James M. and {Rathborne}, J.~M. and {Foster}, J.~B. and {Contreras}, Y. and {Sanhueza}, Patricio and {Stephens}, Ian W. and {Longmore}, S.~N.},
        title = "{MALT90 Kinematic Distances to Dense Molecular Clumps}",
      journal = {\aj},
         year = 2017,
        month = oct,
       volume = {154},
       number = {4},
          eid = {140},
        pages = {140},
          doi = {10.3847/1538-3881/aa86ad},
       adsurl = {https://ui.adsabs.harvard.edu/abs/2017AJ....154..140W}
}

@ARTICLE{bj16,
       author = {{Bjerkeli}, Per and {van der Wiel}, Matthijs H.~D. and {Harsono}, Daniel and {Ramsey}, Jon P. and {J{\o}rgensen}, Jes K.},
        title = "{Resolved images of a protostellar outflow driven by an extended disk wind}",
      journal = {\nat},
         year = 2016,
        month = dec,
       volume = {540},
       number = {7633},
        pages = {406-409},
          doi = {10.1038/nature20600},
archivePrefix = {arXiv},
       eprint = {1612.05148},
 primaryClass = {astro-ph.SR},
       adsurl = {https://ui.adsabs.harvard.edu/abs/2016Natur.540..406B}
}

@ARTICLE{maud2015,
   	author = {{Maud}, L.~T. and {Moore}, T.~J.~T. and {Lumsden}, S.~L. and {Mottram}, J.~C. and {Urquhart}, J.~S. and {Hoare}, M.~G.},
    	title = "{A distance-limited sample of massive molecular outflows}",
  	journal = {\mnras},
     	year = 2015,
    	month = oct,
   	volume = {453},
   	number = {1},
    	pages = {645-665},
      	doi = {10.1093/mnras/stv1635},
archivePrefix = {arXiv},
   	eprint = {1509.00199},
 primaryClass = {astro-ph.SR},
   	adsurl = {https://ui.adsabs.harvard.edu/abs/2015MNRAS.453..645M}
}

@ARTICLE{yang2018,
   	author = {{Yang}, A.~Y. and {Thompson}, M.~A. and {Urquhart}, J.~S. and {Tian}, W.~W.},
    	title = "{Massive Outflows Associated with ATLASGAL Clumps}",
  	journal = {\apjs},
     	year = 2018,
    	month = mar,
   	volume = {235},
   	number = {1},
      	eid = {3},
    	pages = {3},
      	doi = {10.3847/1538-4365/aaa297},
archivePrefix = {arXiv},
   	eprint = {1712.04599},
 primaryClass = {astro-ph.GA},
   	adsurl = {https://ui.adsabs.harvard.edu/abs/2018ApJS..235....3Y}
}

@ARTICLE{li2020,
   	author = {{Li}, Shanghuo and {Sanhueza}, Patricio and {Zhang}, Qizhou and {Nakamura}, Fumitaka and {Lu}, Xing and {Wang}, Junzhi and {Liu}, Tie and {Tatematsu}, Ken'ichi and {Jackson}, James M. and {Silva}, Andrea and {Guzm{\'a}n}, Andr{\'e}s E. and {Sakai}, Takeshi and {Izumi}, Natsuko and {Tafoya}, Daniel and {Li}, Fei and {Contreras}, Yanett and {Morii}, Kaho and {Kim}, Kee-Tae},
    	title = "{The ALMA Survey of 70 {\ensuremath{\mu}}m Dark High-mass Clumps in Early Stages (ASHES). II. Molecular Outflows in the Extreme Early Stages of Protocluster Formation}",
  	journal = {\apj},
     	year = 2020,
    	month = nov,
   	volume = {903},
   	number = {2},
      	eid = {119},
    	pages = {119},
      	doi = {10.3847/1538-4357/abb81f},
archivePrefix = {arXiv},
   	eprint = {2009.05506},
 primaryClass = {astro-ph.GA},
   	adsurl = {https://ui.adsabs.harvard.edu/abs/2020ApJ...903..119L}
}

@ARTICLE{bally2016,
       author = {{Bally}, John},
        title = "{Protostellar Outflows}",
      journal = {\araa},
         year = 2016,
        month = sep,
       volume = {54},
        pages = {491-528},
          doi = {10.1146/annurev-astro-081915-023341},
       adsurl = {https://ui.adsabs.harvard.edu/abs/2016ARA&A..54..491B}
}

@ARTICLE{cyganowski2008,
       author = {{Cyganowski}, C.~J. and {Whitney}, B.~A. and {Holden}, E. and {Braden}, E. and {Brogan}, C.~L. and {Churchwell}, E. and {Indebetouw}, R. and {Watson}, D.~F. and {Babler}, B.~L. and {Benjamin}, R. and {Gomez}, M. and {Meade}, M.~R. and {Povich}, M.~S. and {Robitaille}, T.~P. and {Watson}, C.},
        title = "{A Catalog of Extended Green Objects in the GLIMPSE Survey: A New Sample of Massive Young Stellar Object Outflow Candidates}",
      journal = {\aj},
         year = 2008,
        month = dec,
       volume = {136},
       number = {6},
        pages = {2391-2412},
          doi = {10.1088/0004-6256/136/6/2391},
archivePrefix = {arXiv},
       eprint = {0810.0530},
 primaryClass = {astro-ph},
       adsurl = {https://ui.adsabs.harvard.edu/abs/2008AJ....136.2391C}
}
 
\end{document}